\documentclass[reprint,prb,preprintnumbers,superscriptaddress,amsmath,amssymb]{revtex4-1}
\usepackage{hyperref}
\usepackage{natbib}
\usepackage{graphicx}
\usepackage{dcolumn}
\usepackage{bm}
\usepackage{siunitx}

\begin{document}

\bibliographystyle{unsrtnat}
\title{\large {Intervalley biexcitons and many-body effects in monolayer MoS$_2$}}

\vspace{0.5cm}
\author{Edbert J. Sie}
		\affiliation{Department of Physics, Massachusetts Institute of Technology, Cambridge, MA 02139, USA}
\author{Alex J. Frenzel}
		\affiliation{Department of Physics, Massachusetts Institute of Technology, Cambridge, MA 02139, USA}		
		\affiliation{Department of Physics, Harvard University, Cambridge, MA 02138, USA}
\author{Yi-Hsien Lee}
		\affiliation{Materials Science and Engineering, National Tsing-Hua University, Hsinchu 30013, Taiwan}
\author{Jing Kong}
		\affiliation{Department of Electrical Engineering and Computer Science, Massachusetts Institute of Technology, Cambridge, MA 02139, USA}		
\author{Nuh Gedik}
		\altaffiliation{gedik@mit.edu}
		\affiliation{Department of Physics, Massachusetts Institute of Technology, Cambridge, MA 02139, USA}		
\date{\today}
\vspace{0.5cm}

\begin{abstract}
Interactions between two excitons can result in the formation of bound quasiparticles, known as biexcitons. Their properties are determined by the constituent excitons, with orbital and spin states resembling those of atoms. Monolayer transition metal dichalcogenides (TMDs) present a unique system where excitons acquire a new degree of freedom, the valley pseudospin, from which a novel \textit{intervalley} biexciton can be created. These biexcitons comprise two excitons from different valleys, which are distinct from biexcitons in conventional semiconductors and have no direct analogue in atomic and molecular systems. However, their valley properties are not accessible to traditional transport and optical measurements. Here, we report the observation of intervalley biexcitons in the monolayer TMD MoS$_2$ using ultrafast pump-probe spectroscopy. By applying broadband probe pulses with different helicities, we identify two species of intervalley biexcitons with large binding energies of 60 meV and 40 meV. In addition, we also reveal effects beyond biexcitonic pairwise interactions in which the exciton energy redshifts at increasing exciton densities, indicating the presence of many-body interactions among them.
\end{abstract}
\maketitle

\section{Introduction}
Monolayer transition metal dichalcogenides (TMDs) comprise a new class of atomically thin semiconducting crystals in which electrons exhibit strong spin-valley coupling that results in novel valleytronic properties when interrogated with circularly polarized light \cite{DiXiao12}. These include valley-selective photoexcitation \cite{Mak12,Zeng12,Cao12}, valley Hall effect \cite{Mak14}, valley-tunable magnetic moment \cite{Wu13}, and valley-selective optical Stark effect \cite{Sie14,Kim14}. Unlike traditional semiconductors, the Coulomb interaction in monolayer TMDs is unusually strong because screening is greatly suppressed and spatial overlap of the interaction is much larger. This enhances the stability of a variety of excitonic quasiparticles with extremely large binding energies, including excitons \cite{CommentExciton,Hill15, Klots14,Chernikov14,Ye14,Zhu14,He14,Wang14,Ugeda14}, trions \cite{Mak13,Ross13,Lui14}, and exciton-trion complexes \cite{Singh14}.

In addition to excitons and trions, monolayer TMDs should also host biexcitons, van der Waals quasiparticles formed from two neutral excitons bound by residual Coulomb fields. Moreover, the unique spin-valley coupling of these electrons, which also behave as massive Dirac fermions at two different valleys (Fig 1a), offers an ideal system to form unique intervalley biexcitons \cite{Mai14} that have two-dimensional positronium-molecular-like structure. Apart from the large binding energies, they are also expected to show novel properties such as entanglement between the pair of valley pseudospins. Thorough investigation of these properties is crucial to assess their potential use in applications. Advanced experimental probes are needed to uncover these unique quasiparticles that cannot be accessed by more conventional techniques.

Transient absorption spectroscopy is ideally suited to access intervalley biexciton states via two-step excitation (Fig 1b). In this experiment, an ultrashort laser pulse is split into two portions: the first pulse (pump) is used to create a population of excitons, $\left|0\right\rangle\rightarrow\left|x\right\rangle$, and the second pulse (probe) is used to induce a second transition to form biexcitons, $\left|x\right\rangle\rightarrow\left|xx\right\rangle$ (see Appendix \ref{app:setup}). In monolayer TMDs, there are two degenerate valleys (K and K$^{\prime}$) where different excitons can be created using left ($\sigma^-$) and right ($\sigma^+$) circularly polarized light \cite{DiXiao12,Mak12,Zeng12,Cao12}. In order to form intervalley biexcitons, we used a succession of pump and probe pulses with opposite helicities. The biexcitons are revealed as a pump-induced absorption of the probe pulse at energy slightly below the primary exciton absorption peaks.

\begin{figure*}[t]
	\includegraphics[width=0.77\textwidth]{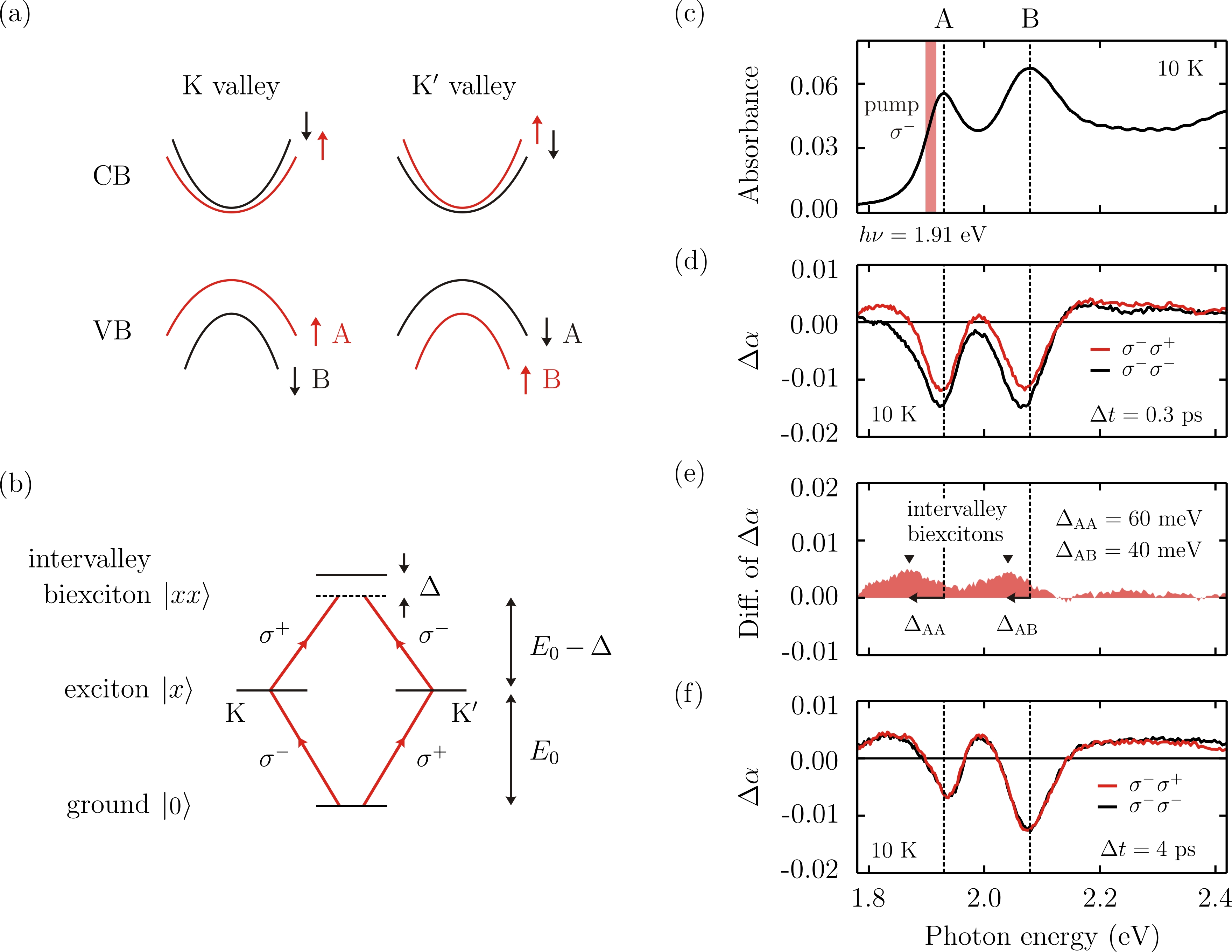}
	\caption{(a) Schematic band structure of the two valleys. (b) Optical transitions to intervalley biexciton state. (c) Measured equilibrium absorbance of monolayer MoS$_2$. (d) Pump-induced absorption of the probe pulses at $\Delta t = 0.3$ ps using different helicities. (e) The difference between the two $\Delta\alpha$ spectra shown in (d), where the biexciton binding energies are obtained. (f) $\Delta\alpha$ spectra at $\Delta t = 4$ ps.}
	\label{fig:Fig1}
\end{figure*}

In this Article, we show that there are two intervalley biexciton species in monolayer TMD MoS$_2$, which we identify as AA biexcitons and AB heterobiexcitons. We measure the binding energies of these biexcitons to be 60 meV and 40 meV, respectively. Experiments using excess pump photon energy reveal the stability of the biexcitons at high temperatures. We also investigate the effect of high excitation densities on the excitons, which shows the presence of many-body effects in monolayer MoS$_2$.

\section{Results and Discussion}
\subsection{Intervalley biexcitons}
The equilibrium absorption spectrum of monolayer MoS$_2$, measured using differential reflectance microscopy at 10 K, consists of two exciton resonances ($E_A = 1.93$ eV and $E_B = 2.08$ eV) and a background from higher-energy states \cite{Klots14} (Fig 1c). To create a population of excitons at the K valley, we used $\sigma^-$ pump pulses with photon energy tuned near the A exciton resonance ($h\nu = 1.91$ eV) and fluence \SI{5}{\micro\J}\text{/cm}$^2$. Further excitation to the intervalley biexciton state can be detected from induced absorption ($\Delta\alpha > 0$) of $\sigma^+$ probe pulses. In an ideal system with negligible scattering, one would also expect to see optical bleaching ($\Delta\alpha < 0$) at the A exciton transition using $\sigma^-$ probe pulses and no bleaching anywhere using $\sigma^+$ probe pulses. Fig 1d shows a pair of $\Delta\alpha$ spectra measured using broadband probe pulses with helicities $\sigma^-$ (black) and $\sigma^+$ (red) at $\Delta t = 0.3$ ps. This time delay was chosen to avoid contaminating effects of coherent light-matter interaction \cite{Sie14, Kim14}. In contrast to what is expected, we observed strong bleaching peaks not only at A but also at B exciton transitions, and strikingly, these two peaks were present in both spectra measured using different helicities \cite{Mai14}.

The unexpected bleaching of the B exciton transition can only originate from electron state filling in the conduction band. This is because the pump photon energy is insufficient to excite holes to form a B exciton, and hole scattering between the A and B bands is very unlikely due to the large energy splitting (150 meV). Meanwhile, the photoexcited electron spin-up state (for A exciton) at the K valley can exhibit an \textit{intravalley} spin reversal to occupy the electron spin-down state (for B exciton). This process can be mediated by flexural phonons \cite{Song13}, and it can occur during the pump pulse duration (Appendix \ref{app:spin_reversal}). On the other hand, the bleaching peaks that are observed using the opposite probe helicity can be explained by \textit{intervalley} scattering, which is expected to be very fast due to electron-hole exchange interaction in this material \cite{Mai14,Yu14, ZhuCR14}. In particular, valley excitons with finite in-plane momentum can have exchange interaction that generates an in-plane effective magnetic field, around which the exciton pseudospin precesses from K to K$^{\prime}$ valley. Calculations show that such valley depolarization can be as fast as tens to hundreds of fs \cite{Yu14}, consistent with our observation.

\begin{figure}[t]
	\includegraphics[width=0.41\textwidth]{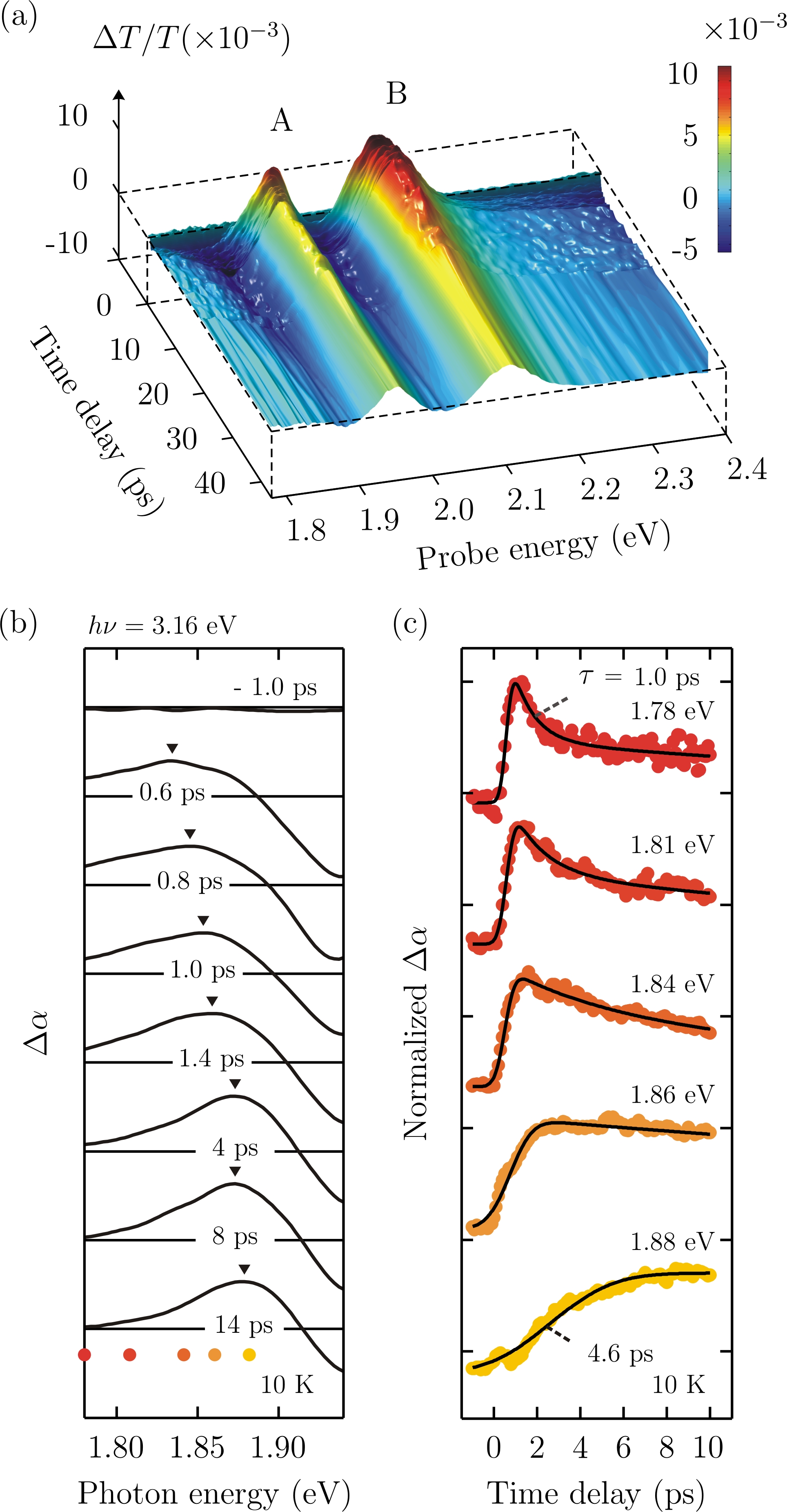}
	\caption{(a) $\Delta T/T$ spectra as a function of $\Delta t$ after pump pulse excitation at $h\nu = 3.16$ eV. (b) $\Delta\alpha$ spectra at increasing time delays, showing the peak sharpening and the shifting to higher probing energy. (c) Time-traces of $\Delta\alpha$ at probe energies as indicated by the colored dots in (b).}
	\label{fig:Fig2abcd}
\end{figure}

\begin{figure}[t]
	\includegraphics[width=0.41\textwidth]{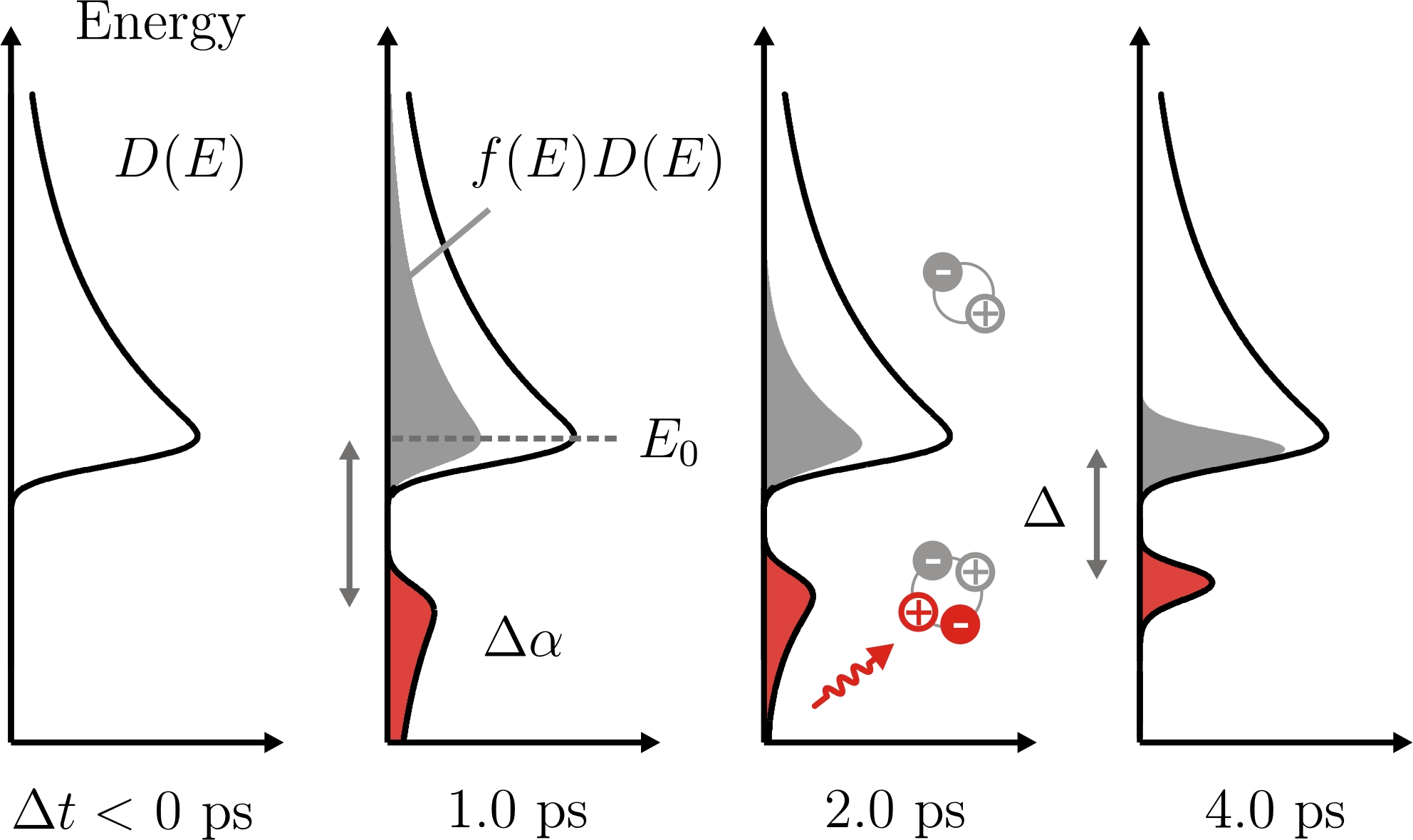}
	\caption{Time evolution of the exciton energy distribution (grey) and the corresponding biexciton induced absorption $\Delta\alpha$ (red) through the cooling process.}
	\label{fig:Fig3abc}
\end{figure}

Comparison of the two spectra in Fig 1d shows that the red curve gains an appreciable offset in certain spectral ranges with respect to the black curve that, according to the pump-probe scheme in Fig 1b, can be attributed to induced absorption to intervalley biexciton states. To isolate the biexciton contribution, we evaluate the difference between the two spectra (Fig 1e) from which we extract the biexciton binding energies ($\Delta_{xx}$). We find peak energies of $\Delta_{\text{AA}} = 60$ meV and $\Delta_{\text{AB}} = 40$ meV, consistent with recently estimated values \cite{Mai14,Shang15}. Since we only photoexcited the A excitons at resonant excitation, we expect the biexcitons to contain at least one A exciton. Thus, by the proximity of the biexciton peaks in Fig 1e to the exciton resonances, we deduce that the former is comprised of two A excitons (AA biexcitons), while the latter consists of one A and one B exciton (AB heterobiexcitons). The intervalley nature of these biexcitons is guaranteed from our measurement protocol, where we used pump and probe pulses with opposite helicities. Additional contributions apart from the intervalley biexcitons are eliminated by taking the difference in $\Delta\alpha$ spectra as shown in Fig 1e. At later time delays, both $\Delta\alpha$ spectra become nearly identical (Fig 1f) because intervalley scattering establishes balance between the exciton populations at the two valleys. Biexciton formation can then be induced from these excitons via absorption of probe pulses with either polarization ($\sigma^-$ and $\sigma^+$).

We measured biexciton binding energies to be greater than the thermal energy at room temperature, $\Delta_{xx} > 25$ meV. This is expected because the excitons in monolayer MoS$_2$ have large binding energies ($E_b$), with measured values reported from 440 meV \cite{Hill15} to 570 meV \cite{Klots14}. Despite this variation in the reported values, the obtained binding energies are consistent with theoretical models \cite{Kleinman83,Singh96} that predict $\Delta_{\text{AA}} =$ (0.13$-$0.23) $E_b$ in monolayer MoS$_2$ \cite{Thilagam14}. The large biexciton binding energies share the same origin as those of the excitons where, in the 2D limit, quantum confinement and suppressed screening greatly enhance the Coulomb interaction in this system.

\subsection{Cooling process}
We now turn to discuss the time evolution of biexciton formation upon photoexcitation using excess pump photon energy ($h\nu = 3.16$ eV) and exciton excitation density of $1.4 \times 10^{12}$ cm$^{-2}$. In this section, we measured the differential transmittance $\Delta T/T$ as a function of probe photon energy and time delay (Fig 2a). We plot the corresponding $\Delta\alpha$ spectra at different time delays in Fig 2b, expanded around the A exciton resonance. The induced absorption peak exhibits a gradual shift in energy from 1.83 eV at 0.6 ps to 1.88 eV at 14 ps, which is accompanied by a peak sharpening. To elucidate this behavior we plot, in Fig 2c, the $\Delta\alpha$ time-traces (normalized) between energies 1.78$-$1.88 eV, as indicated by the colored dots in Fig 2b. We find that the decay at 1.78 eV is accompanied by a buildup at 1.88 eV. The latter value is consistent with the peak position of the biexciton signature measured in the near resonant excitation.

We interpret this observation as resulting from the exciton cooling process, where the energy distribution of the photoexcited excitons varies with the time delay, as depicted in Fig 3. In the first panel, we show the density of exciton states $D(E)$ surrounding the A exciton resonance. In our experiment, the non-resonant excitation by the pump pulse ($h\nu > E_\text{A}$) imparts an excess energy of $\delta E \sim 1$ eV per exciton. This leads to the immediate formation of a hot exciton gas ($T_{\text{e}} \sim 10^3$ K), where the population of highly energetic excitons $f(E)D(E)$ at $E > E_\text{A}$ becomes substantial. This is shown on the second panel of Fig 3 (grey), including the associated $\Delta\alpha$ spectra (red). The biexciton energy can now be expressed as
\begin{equation}
E_{\text{AA}} = (E_\text{A} + \delta E)_{\text{pump}} + (E_\text{A} - \Delta - \delta E)_{\text{probe}}
\end{equation}
where $\delta E$ is the excess exciton energy whose distribution depends on $T_\text{e}$. Hence, we only need to create a probe exciton with energy lower than $E_\text{A}-\Delta$ to form a biexciton of energy $E_{\text{AA}}$. This explains the low-energy tail of $\Delta\alpha$ at $\Delta t \leq 1.4$ ps (Fig 2b). Moreover, the presence of biexciton absorption at such an early relaxation stage of the hot exciton gas shows that they are stable at high temperatures. At later time delays (Fig 3, right panels), the highly energetic excitons gradually relax into the lowest state, via releasing energy to the lattice and the substrate, to form a cold exciton gas. During this process, the $\Delta\alpha$ spectral weight (red) at lower energy gradually climbs to higher energy, consistent with the observed dynamics in Fig 2c and the $\Delta\alpha$ peak shifting and sharpening in Fig 2b.

\begin{figure}[t]
	\includegraphics[width=0.41\textwidth]{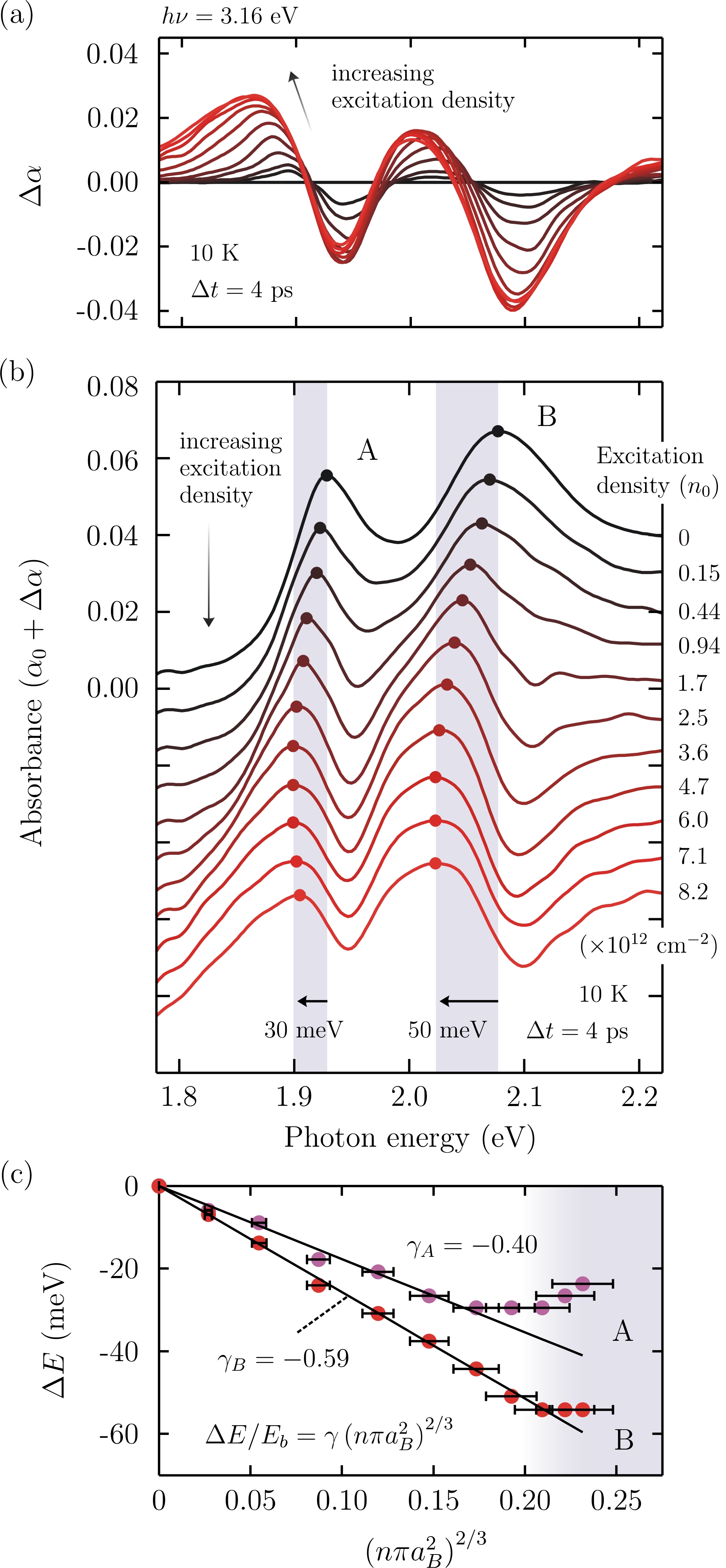}
	\caption{(a) Induced absorption spectra at increasing excitation densities of up to $8.2 \times 10^{12}$ cm$^{-2}$. (b) Transient absorption spectra as extracted from (a), showing the red-shift of the exciton peaks due to many-body effects. (c) Energy shift at densities $n(\Delta t = 4\ \text{ps})$, scaled by the exciton size.}
	\label{fig:Fig4}
\end{figure}

\subsection{Many-body effects}
We have shown that the biexciton binding energies in monolayer MoS$_2$ are rather large, rendering biexcitons stable against thermal disintegration above room temperature. This finding results from the strong, long-range Coulomb interaction in this material, which also causes many-body effects to play a significant role at higher excitation densities. The interactions introduced by photoexcited carriers often result in an exciton energy shift \cite{Boldt85,Zimmermann88,Schmitt85,Schmitt89,Steinhoff14}, which can be used as a sensitive indicator of many-body correlations. To investigate these effects, we measured a series of $\Delta\alpha$ spectra at increasing excitation densities using pump photon energy 3.16 eV (Fig 4a). The 4 ps time delay was chosen to guarantee that all excitations relaxed to the lowest excited state (A exciton) and that these excitons reached a quasi-equilibrium temperature shared by the lattice and the substrate. These $\Delta\alpha$ spectra were added to the measured equilibrium absorption spectrum $\alpha_0$, yielding the transient absorption spectra shown in Fig 4b. As the excitation density is increased, the exciton absorption peak shifts to lower energies and develops a low-energy shoulder. The shoulder corresponds to biexciton formation, as discussed above, while the peak-shift results from many-body interactions, consistent with a recent theoretical study \cite{Steinhoff14}. These peak-shifts can be as large as 30 meV (A) and 50 meV (B) at a density of $4.7 \times 10^{12}$ cm$^{-2}$, with no additional shift up to $8.2 \times 10^{12}$ cm$^{-2}$.

In general, the energy shift is a result of complex many-body interactions, a well-known problem in condensed matter physics that requires synergistic investigation from first-principle calculations and experiments. As anticipated in highly excited semiconductors, photoexcited carriers introduce additional screening and modify exchange-correlation energies \cite{Boldt85, Zimmermann88, Schmitt85,Schmitt89,Steinhoff14}. Renormalization of the electronic structure gives rise to the reduction of both the band gap and the exciton binding energy. Incomplete compensation of the two effects often results in a net energy shift of the exciton peak. The behavior, at which the energy shift varies as a function of exciton density, can be studied to determine the detailed contributions from different interaction terms. We examined our results by rescaling the energy shift ($\Delta E$) and the time-delayed density ($n$) into dimensionless parameters (Appendix \ref{app:manybody}). We found that in the low-density regime they follow a simple formula (Fig 4c)
\begin{equation}
\frac{\Delta E}{E_b} = \gamma \left(n\pi a_B^2\right)^{k}
\end{equation}
where $E_b$ ($= 440$ meV \cite{Hill15}) is the exciton binding energy, $a_B$ ($= 1$ nm \cite{Steinhoff14}) is the exciton Bohr radius, and $\gamma$ is a dimensionless factor. We found that the experimental results are best fitted with $k = 2/3$, while fits using other exponents are relatively poor (Appendix \ref{app:manybody}). By fitting the first seven data points with $k = 2/3$, we obtained $\gamma_A = -0.40 \pm 0.02$ and $\gamma_B = -0.59 \pm 0.01$, while the remaining data points deviate from linear behavior. Equation (2) shows that $\Delta E$ obeys a power-law behavior with density, until the inter-exciton distance ($1/\sqrt{n}$) approaches the exciton Bohr radius, where the Mott transition is expected \cite{Zimmermann88}. In this extreme situation, the electron is in a frustrated state where it cannot distinguish which hole belongs to its exciton pair, thus the system has a tendency to become metallic.

This power-law behavior at low-density regime can be perceived in a physically intuitive picture from its exponent. We note that the measured energy shift ($\Delta E < 0$) is the energy gain for each exciton in the presence of collective photoexcited excitons, as compared to the energy of an isolated exciton. If $\Delta E \propto \sqrt{n}$, the energy gain is inversely proportional to the mean inter-exciton distance ($1/\sqrt{n}$) where short-range interaction dominates. If $\Delta E \propto n$, the energy gain is proportional to the number of surrounding excitons per unit area ($n$), where long-range interaction dominates. These two scenarios have subtle differences in determining the major contributions to the energy shift, for which first-principle calculations can be helpful \cite{Steinhoff14}. The obtained relation $\Delta E \propto n^{2/3}$ shows that the exciton interactions in monolayer TMDs fall in between the two scenarios where both short- and long-range interactions are comparable, even in the low-density regime.

\section{Conclusions}
By using transient absorption spectroscopy we have observed intervalley biexcitons in monolayer MoS$_2$, measured their binding energies, and monitored their relaxation processes. Studying intervalley biexcitons could offer a new concept of quasiparticles in solids where valley-pseudospin states play significant roles, apart from the usual atom-like orbital and spin states. The large biexciton binding energies in this material offer a promising direction to search for higher-order bound excitons such as triexcitons and the elusive quadexcitons \cite{Stone09, Turner10} as well as their interplay with spin-valley degrees of freedom. We also found that, while two excitons can interact and form a biexciton, the interactions between \textit{free} excitons in this system reveal a large energy red-shift that obeys a simple power-law behavior with density. This experimentally obtained relation between the energy shift and the density, as well as the obtained biexciton binding energies, should be used to aid future theoretical works and additional experiments in exploring many-body physics in atomically thin materials.
\newline
\indent
\textit{Note added}: During the review process, a related work was reported by another group \cite{You15}.

\begin{acknowledgments}
The authors acknowledge technical assistance by Q. Ma and Y. Bie during the measurement of the equilibrium absorption of monolayer MoS$_2$, and helpful discussions with C. H. Lui, Z. Alpichshev, T. Rohwer, C. Gies, F. Mahmood and H. O. H. Churchill. This work is supported by U.S. Department of Energy (DOE) award numbers DE-FG02-08ER46521 and DE-SC0006423 (data acquisition and analysis). Y.-H.L. and J.K. acknowledge support from NSF DMR 0845358 (material growth and characterization). Y.-H.L. also acknowledges partial support from the Ministry of Science and Technology of the Republic of China (103-2112-M-007-001-MY3 \& 103-2633-M-007-001).
\end{acknowledgments}

\appendix
\section{Transient absorption spectroscopy}
\label{app:setup}
In our experiments, we used a Ti:sapphire regenerative amplifier producing laser pulses at 30 kHz, with center wavelength 785 nm and duration 50 fs FWHM. Each pulse was split into two arms. For the pump arm, the pulses were frequency-converted using an optical parametric amplifier (for resonant excitation) or a second-harmonic crystal (for non-resonant excitation), and then chopped at 7.5 kHz. For the probe arm, the pulses were sent through a delay stage and a white-light continuum generator ($h\nu =$ 1.78$-$2.48 eV, chirp-corrected). The two beams were focused onto the sample with \SI{450}{\micro\meter} (pump) and \SI{150}{\micro\meter} (probe) FWHM diameters. The probe beam was reflected or transmitted from the sample to a monochromator with FWHM resolution 1 nm and a photodiode for detection. Lock-in detection at 7.5 kHz allowed measurement of fractional changes in reflectance $\Delta R/R$ or transmittance $\Delta T/T$ as small as  $10^{-4}$. By scanning the grating and the delay stage, we were able to measure $\Delta R/R$ or $\Delta T/T$ as a function of energy and time delay $\Delta t$, from which the induced absorptance $\Delta\alpha$ was obtained using Kramers-Kronig analysis (Appendix \ref{app:kramers}). The pump fluence was varied by a combination of a half-wave plate and polarizer, allowing us to tune the exciton excitation density. High-quality monolayers of MoS$_2$ were CVD-grown on a sapphire substrate \cite{Lee12,Lee13} and mounted inside a cold-finger cryostat with temperature of 10 K for all measurements in this study.

\section{Kramers-Kronig analysis}
\label{app:kramers}
Pump-probe experiments detect small changes in probe reflectance (or transmittance) that is induced by pump excitation. This gives the differential reflectance $\Delta R/R$ as a function of energy and time delay, from which we can obtain the transient reflectance, $R(t) = R_0 (1 + \Delta R(t)/R_0)$, where $R_0$ is the reflectance of the system in equilibrium. In fact, the absorptance $\alpha$ (or the induced absorptance $\Delta\alpha$) is what we really want because it provides the explicit information about the optical transition matrix element of the system. The absorptance and the reflectance are related through the complex dielectric function $\tilde{\epsilon}$; this relation can be derived using Maxwell equations \cite{Falkovsky08,Stauber08}. We obtain $\tilde{\epsilon}(\omega,t)$ by fitting $R(\omega,t)$ using a Kramers-Kronig (KK) constrained variational analysis \cite{Kuzmenko05}. Finally, we construct $\alpha(\omega,t)$ by repeating this procedure at different time delays. The details of the above procedure are described as follows.

First, we want to find the relation between the complex dielectric function and the optical properties such as reflectance, transmittance and absorptance by using Maxwell's equations. Note that it is important to include the substrate influence on electromagnetic radiation especially for atomically-thin materials \cite{Falkovsky08,Stauber08}. Here, the current density in a monolayer MoS$_2$ sample is described by a delta function, $j_x = \tilde{\sigma}(\omega)\delta(z)E_x$ where $\tilde{\sigma}$ is the complex conductivity and $E_x$ is the $x$-component of the probe electric field (along the sample's surface). By substituting this into Maxwell's equations and using the appropriate boundary conditions between the monolayer and the substrate, we can obtain the reflectance as
\begin{equation}
R(\omega) = \frac{(1-n_s-\frac{\omega d}{c}\epsilon_2)^2+(\frac{\omega d}{c}(\epsilon_1 - 1))^2}{(1+n_s+\frac{\omega d}{c}\epsilon_2)^2+(\frac{\omega d}{c}(\epsilon_1 - 1))^2}
\label{eqnR}
\end{equation}
and the transmittance as
\begin{equation}
T(\omega) = \frac{4n_s}{(1+n_s+\frac{\omega d}{c}\epsilon_2)^2+(\frac{\omega d}{c}(\epsilon_1 - 1))^2}
\end{equation}
where $n_s$ is the substrate's refractive index (1.7675 for sapphire at photon energy of 2.07 eV), $d$ is the effective thickness of the monolayer (0.67 nm \cite{Mak13}), $\epsilon_1$ and $\epsilon_2$ are the real and imaginary parts of the dielectric function, respectively. Here, the 2D dielectric function is expressed as
\begin{equation}
\tilde{\epsilon}(\omega) = 1+ \frac{4\pi i \tilde{\sigma}/d}{\omega}
\end{equation}
Meanwhile, the absorptance can be expressed as
\begin{equation}
\alpha(\omega) = \frac{4\frac{\omega d}{c}\epsilon_2}{(1+n_s+\frac{\omega d}{c}\epsilon_2)^2+(\frac{\omega d}{c}(\epsilon_1 - 1))^2}
\label{eqnA}
\end{equation}
These expressions are exact, and they are valid for any monolayer materials on a dielectric substrate. We find that the presence of the substrate significantly influences the optical properties of the monolayer MoS$_2$ above it. As compared to an isolated monolayer MoS$_2$, the reflectance is enhanced, while both the transmittance and the absorptance are reduced. In graphene, the above expressions can be further simplified because the real part of its dielectric function is featureless in the visible spectrum ($\epsilon_1 \sim 1$, negligible $\sigma_2$ \cite{Li08}). This is, however, not the case for monolayer MoS$_2$, and we must include both the real and imaginary parts of the dielectric function to obtain accurate results.

In our analysis, we used the equilibrium absorptance $\alpha$ of monolayer MoS$_2$ that is measured using differential reflectance microscopy (Fig 1c, main text). The absorptance spectrum contains peaks from the A exciton at 1.93 eV and from the B exciton at 2.08 eV. The equilibrium reflectance $R_0$ can then be constructed from $\alpha$ by finding the appropriate complex dielectric function $\tilde{\epsilon}$ as expressed in equations (\ref{eqnR}) and (\ref{eqnA}). To do this, we implemented a Kramers-Kronig (KK) constrained variational analysis \cite{Kuzmenko05} to extract $\tilde{\epsilon}$ from the measured $\alpha$. Here, the total dielectric function is constructed by many Drude-Lorentz oscillators, which are anchored at equidistant energy spacing, in the following form
\begin{equation}
\tilde{\epsilon}(\omega) = \epsilon_{\infty} + \sum_{k=1}^N \frac{\omega_{p,k}^2}{\omega_{0,k}^2 - \omega^2 - i\omega \gamma_k}
\end{equation}
In our calculations, we used $N = 40$ oscillators with a fixed linewidth of $\gamma_k = 50$ meV spanning the energy range of 1.77 eV $\leq\omega_{0,k}\leq$ 2.40 eV, and we found that these parameters can fit the absorptance spectrum well. We can then construct $R_0$ spectrum by using $\tilde{\epsilon}$ obtained from the above analysis.

The transient absorptance spectra $\alpha(t)$ can be obtained by performing similar (KK) analysis. This time we inferred the absorptance from the reflectance at different time delays: $R(t) = R_0 (1+\Delta R(t)/R_0)$, where the differential reflectance $\Delta R(t)/R_0$ is measured directly from the experiments. Similar procedure also applies to the transmission geometry.

\section{Intravalley electron spin reversal}
\label{app:spin_reversal}
In order to understand the bleaching of B exciton transition, we note that at the K valley the electron states in the conduction band are split by the spin-orbit coupling into two spin states, where the electron spin-up state (for A exciton) is 3 meV lower in energy than the electron spin-down state (for B exciton) \cite{Kosmider13, Liu13, Kormanyos13, Cheiw12}. These two states preserve the good spin quantum numbers (out of plane) due to the $\sigma_{\text{h}}$ mirror symmetry of the lattice. Resonant excitation of the A exciton using $\sigma^-$ pump pulses only populates the electron spin-up state at K valley (see Fig 1a), and should in principle not bleach the B exciton transition. Hence, the observed bleaching of B exciton can only be explained by the electron spin reversal in the conduction band that immediately occurs during the pump pulse duration (160 fs). In this flexible material, intravalley electron spin reversal can be mediated by flexural phonons in the realm of Elliott-Yafet spin-flip mechanism \cite{Song13}. Spin-flip transition is allowed in the scattering via long-wavelength in-plane optical phonon and out-of-plane acoustic phonon, for which the respective deformation potentials are even and odd with respect to mirror symmetry. Because of the small spin-splitting in the conduction band, the strong spin-orbit coupling and the flexible lattice structure, the spin lifetime could be as short as 50 fs for suspended monolayer MoS$_2$ at room temperature. Supporting the membrane with sapphire substrate at 10 K can in principle prolong the spin lifetime \cite{Song13} but substrate roughness, domain boundaries and impurities could conversely increase the rate of carrier collisions that will enhance the spin-reversal scattering rate. The fast electron spin reversal could thus explain the B exciton bleaching shown in Fig 1d.

\section{Estimating the excitation density from the measured pump fluence}
\label{app:density}
The pump and probe pulses were focused at the sample to \SI{450}{\micro\meter} and \SI{150}{\micro\meter} FWHM, respectively, within which 50\% of the energy per pulse is contained. We are interested in finding the effective pump fluence that is sampled by the probe pulse, with area of only one-third of the pump diameter (7.5\% of the energy). For instance, a total pump power of 1 mW (30 kHz) measured at the sample position gives an effective pump fluence of \SI{14}{\micro\J}/cm$^2$ within the probed area. The interference from the substrate further reduces the pump intensity to
\begin{equation}
\frac{I}{I_0} = \left|1+r\right|^2 = \frac{4}{\left(1+n_s\right)^2} = 0.52
\end{equation}
where the refractive index of the sapphire substrate at 3.16 eV photon energy is $n_s = 1.7865$. The absorptance of suspended monolayer MoS$_2$ at 3.16 eV photon energy was measured to be around 25\% \cite{Mak10}. Finally, by taking all of the above considerations into account, we can estimate the e-h excitation density as
\begin{eqnarray}
\begin{split}
n_0 &= \frac{\SI{14}{\micro\J}\text{/cm}^2\times 0.52 \times 0.25}{3.16\ \text{eV}} \\
  &= \frac{\SI{1.8}{\micro\J}\text{/cm}^2}{3.16\ \text{eV}} = 3.6 \times 10^{12}\ \text{cm}^{-2}
\end{split}
\end{eqnarray}
Pump intensity variations due to the spatially-varying (gaussian) profile and the pulse-to-pulse power fluctuation contribute to the fluence uncertainty. With these considerations, we estimated the fluence variation of less than $\pm 5\%$ in our experiments.

\section{Time-delayed density and energy shift power-law behavior with density}
\label{app:manybody}
In order to obtain an accurate behavior of the energy shift, the exciton density must be carefully determined. We note that the absorptance spectra were measured at time delay of 4 ps after photoexcitation. This means the actual exciton density ($n$) may be slightly smaller than the excitation density ($n_0$). During this process, exciton-exciton annihilation could happen with a decay rate that depends on the initial excitation density, as studied by Sun \textit{et al.} \cite{Sun14} and Kumar \textit{et al.} \cite{Kumar14}. We incorporated this process in our analysis, and used the obtained time-delayed density $n(t)$ in our main text
\begin{eqnarray}
n(t)=\frac{n_0}{\left(1+k_A n_0 t\right)}
\end{eqnarray}
where $n_0$ is the initial photoexcited exciton density, $k_A$ ($=(4.3 \pm 1.1) \times 10^{-2}\  \text{cm}^2/\text{s}$, \cite{Sun14}) is the exciton-exciton annihilation rate, and $t$ ($= 4$ ps) is the time-delay after photoexcitation. In order to get a better physical picture, we scaled the density $n(t)$ with the exciton Bohr radius $a_B$ into a dimensionless density $n\pi a_B^2$. In this way, we can think of the excitons as rigid bodies and determine the critical density ($n \pi a_B^2 \sim 1$) at which the excitons are closely packed. 

\begin{figure}[t]
\renewcommand{\thefigure}{E1}
	\includegraphics[width=0.48\textwidth]{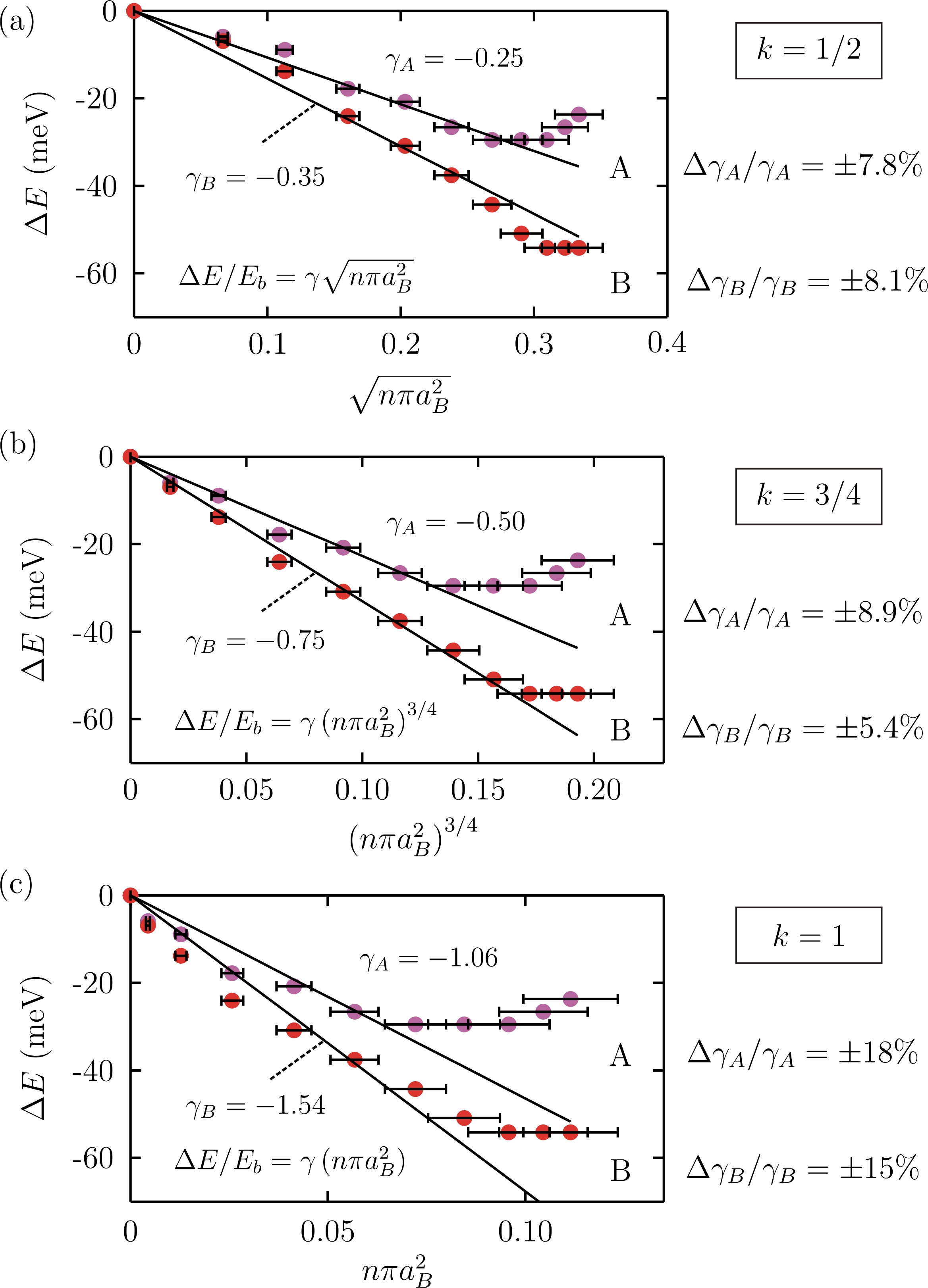}
	\caption{Energy shift vs dimensionless density at (a) $k = 1/2$, (b) $k = 3/4$  and (c) $k = 1$.}
	\label{fig:FigS1abc}
\end{figure}

We plot the power-law dependence of the energy shift $\Delta E$ as a function of dimensionless density
\begin{eqnarray}
\Delta E/E_b = \gamma \left(n\pi a_B^2\right)^k
\end{eqnarray}
where $E_b$ ($= 440$ meV, \cite{Hill15}) is the exciton binding energy, $\gamma$ is a dimensionless factor, and $k$ is the exponent. We also performed an uncertainty analysis to determine the horizontal error-bars of every data points using the quadratic-sum method which yields a fractional uncertainty of
\begin{eqnarray}
\frac{\Delta(n\pi a_B^2)^k}{(n\pi a_B^2)^k} = k\frac{\sqrt{\eta^2+(\Delta k_A n_0 t)^2}}{1+k_A n_0 t}
\end{eqnarray}
where $\eta$ ($=5\%$) is the laser fluence fluctuation and $\Delta k_A$ ($=1.1 \times 10^{-2}\  \text{cm}^2/\text{s}$, \cite{Sun14}) is the standard deviation of the exciton-exciton annihilation rate.

In order to determine the correct power-law behavior, we plot the results with three different exponents of $k = 1/2, 3/4, 1$ and make a comparison through a linear fitting of the first seven data points (Fig E1). By using least-square method to fit equation (9) with the measured data, we can obtain their fractional standard deviations ($\Delta\gamma/\gamma$). We found that the best fit can be obtained using $k = 2/3$ (Fig 4c, main text).


\begin{thebibliography}{}
\bibitem{DiXiao12} 	D. Xiao, G. B. Liu, W. Feng, X. Xu, and W. Yao, Phys. Rev. Lett. 108, 196802 (2012).
\bibitem{Mak12}		K. F. Mak, K. He, J. Shan, and T. F. Heinz, Nature Nano. 7, 494 (2012).
\bibitem{Zeng12}	H. Zeng, J. Dai, W. Yao, D. Xiao, and X. Cui, Nature Nano. 7, 490 (2012).
\bibitem{Cao12}		T. Cao, G. Wang, W. Han, H. Ye, C. Zhu, J. Shi, Q. Niu, P. Tan, E. Wang, B. Liu, and J. Feng, Nature Comms. 3, 887 (2012).
\bibitem{Mak14}		K. F. Mak, K. L. McGill, J. Park, and P. L. McEuen, Science 344, 1489 (2014).
\bibitem{Wu13}		S. Wu, J. S. Ross, G. -B. Liu, G. Aivazian, A. Jones, Z. Fei, W. Zhu, D. Xiao, W. Yao, D. Cobden, and X. Xu, Nature Phys. 9, 149 (2013).
\bibitem{Sie14}		E. J. Sie, J. W. McIver, Y. -H. Lee, L. Fu, J. Kong, and N. Gedik, Nature Mater. 14, 290 (2015).
\bibitem{Kim14}		J. Kim, X. Hong, C. Jin, S. -F. Shi, C. -Y. S. Chang, M. -H. Chiu, L. -J. Li, and F. Wang, Science 346, 1205 (2014).
\bibitem{CommentExciton}  An exciton is a bound electron-hole pair, formed by Coulomb attraction between the two charge carriers. The exciton binding energies were measured with reported values ranging from 440 meV \cite{Hill15} to 570 meV \cite{Klots14} for MoS$_2$, 320 meV \cite{Chernikov14, Hill15} to 700 meV \cite{Ye14, Zhu14} for WS$_2$, 370 meV \cite{He14} to 600 meV \cite{Wang14} for WSe$_2$, and 550 meV for MoSe$_2$ \cite{Ugeda14}. The variation of the reported values, even for the same material, could be attributed to the different substrates, procedures of material preparation, and methods of measurement.
\bibitem{Hill15}  H. M. Hill, A. F. Rigosi, C. Roquelet, A. Chernikov, T. C. Berkelbach, D. R. Reichman, M. S. Hybertsen, L. E. Brus, and T. F. Heinz, Nano Lett. 15, 2992 (2015).
\bibitem{Klots14}  A. R. Klots, A. K. M. Newaz, B. Wang, D. Prasai, H. Krzyzanowska, J. Lin, D. Caudel, N. J. Ghimire, J. Yan, B. L. Ivanov, K. A. Velizhanin, A. Burger, D. G. Mandrus, N. H. Tolk, S. T. Pantelides, and K. I. Bolotin, Sci. Rep. 4, 6608 (2014).
\bibitem{Chernikov14} A. Chernikov, T. C. Berkelbach, H. M. Hill, A. Rigosi, Y. Li, O. B. Aslan, D. R. Reichman, M. S. Hybertsen, and T. F. Heinz, Phys. Rev. Lett. 113, 076802 (2014).
\bibitem{Ye14}	Z. Ye, T. Cao, K. O'Brien, H. Zhu, X. Yin, Y. Wang, S. G. Louie, and X. Zhang, Nature 513, 214 (2014).
\bibitem{Zhu14}		B. Zhu, X. Chen, and X. Cui, Sci. Rep. 5, 9218 (2015).
\bibitem{He14}	K. He, N. Kumar, L. Zhao, Z. Wang, K. F. Mak, H. Zhao, and J. Shan, Phys. Rev. Lett. 113, 026803 (2014).
\bibitem{Wang14}	G. Wang, X. Marie, I. Gerber, T. Amand, D. Lagarde, L. Bouet, M. Vidal, A. Balocchi, and B. Urbaszek, Phys. Rev. Lett. 114, 097403 (2015).
\bibitem{Ugeda14}	M. M. Ugeda, A. J. Bradley, S. -F. Shi, F. H. da Jordana, Y. Zhang, D. Y. Qiu, W. Ruan, S. -K. Mo, Z. Hussain, Z. -X. Shen, F. Wang, S. G. Louie, and M. F. Crommie, Nature Mater. 13, 1091 (2014).
\bibitem{Mak13}		K. F. Mak, K. He, C. Lee, G. H. Lee, J. Hone, T. F. Heinz, and J. Shan, Nature Mater. 12, 207 (2013).
\bibitem{Ross13}	J. S. Ross, S. Wu, H. Yu, N. J. Ghimire, A. M. Jones, G. Aivazian, J. Yan, D. G. Mandrus, D. Xiao, W. Yao, and X. Xu, Nature Comms. 4, 1474 (2013).
\bibitem{Lui14}	C. H. Lui, A. J. Frenzel, D. V. Pilon, Y. -H. Lee, X. Ling, G. M. Akselrod, J. Kong, and N. Gedik, Phys. Rev. Lett. 113, 166801 (2014).
\bibitem{Singh14}	A. Singh, G. Moody, S. Wu, Y. Wu, N. J. Ghimire, J. Yan, D. G. Mandrus, X. Xu, and X. Li, Phys. Rev. Lett. 112, 216804 (2014).
\bibitem{Mai14}		C. Mai, A. Barrette, Y. Yu, Y. G. Semenov, K. W. Kim, L. Cao, and K. Gundogdu, Nano Lett. 14, 202 (2014).
\bibitem{Song13}  Y. Song and H. Dery, Phys. Rev. Lett. 111, 026601 (2013).
\bibitem{Yu14}	T. Yu and M. W. Wu, Phys. Rev. B 89, 205303 (2014).
\bibitem{ZhuCR14}	C. R. Zhu, K. Zhang, M. Glazov, B. Urbaszek, T. Amand, Z. W. Ji, B. L. Liu, and X. Marie, Phys. Rev. B 90, 161302 (2014).
\bibitem{Shang15}  J. Shang, X. Shen, C. Cong, N. Peimyoo, B. Cao, M. Eginligil, and T. Yu, ACS Nano 9, 647 (2015).
\bibitem{Kleinman83} 	D. A. Kleinman, Phys. Rev. B 28, 871 (1983).
\bibitem{Singh96}	J. Singh, D. Birkedal, V. G. Lyssenko, and J. M. Hvam, Phys. Rev. B 53, 15909 (1996).
\bibitem{Thilagam14}	A. Thilagam, J. Appl. Phys. 116, 053523 (2014).
\bibitem{Boldt85} F. Boldt, K. Henneberger, and V. May, phys. stat. sol. (b) 130, 675 (1985).
\bibitem{Zimmermann88}	R. Zimmermann, phys. stat. sol. (b) 146, 371 (1988).
\bibitem{Schmitt85}	S. Schmitt-Rink, D. S. Chemla, and D. A. B. Miller, Phys. Rev. B 32, 6601 (1985).
\bibitem{Schmitt89}  S. Schmitt-Rink, D. S. Chemla, and D. A. B. Miller, Adv. Phys. 38, 89 (1989)
\bibitem{Steinhoff14}	A. Steinhoff, M. Rosner, F. Jahnke, T. O. Wehling, and C. Gies, Nano Lett. 14, 3743 (2014).
\bibitem{Stone09} 	K. W. Stone, K. Gundogdu, D. B. Turner, X. Li, S. T. Cundiff, and K. A. Nelson, Science 324, 1169 (2009).
\bibitem{Turner10}		D. B. Turner and K. A. Nelson, Nature 466, 1089 (2010).
\bibitem{You15}  Y. You, X. -X. Zhang, T. C. Berkelbach, M. S. Hybertsen, D. R. Reichman, and T. F. Heinz, Nature Phys. 11, 477 (2015).
\bibitem{Lee12} 	Y. -H. Lee, X. -Q. Zhang, W. Zhang, M. -T. Chang, C. -T. Lin, K. -D. Chang, Y. -C. Yu, J. T. -W. Wang, C. -S. Chang, L. -J. Li, and T. -W. Lin, Adv. Mat. 24, 2320 (2012).
\bibitem{Lee13}		Y. -H. Lee, L. Yu, H. Wang, W. Fang, X. Ling, Y. Shi, C. -T. Lin, J. -K. Huang, M. -T. Chang, C. -S. Chang, M. Dresselhaus, T. Palacios, L. -J. Li, and J. Kong, Nano Lett. 13, 1852 (2013).
\bibitem{Falkovsky08} L. A. Falkovsky, J. Phys.: Conf. Ser. 129, 012004 (2008).
\bibitem{Stauber08}	T. Stauber, N. M. R. Peres, and A. K. Geim, Phys. Rev. B 78, 085432 (2008).
\bibitem{Kuzmenko05} A. B. Kuzmenko, Rev. Sci. Instrum. 76, 083108 (2005).
\bibitem{Li08}  Z. Q. Li, E. A. Henriksen, Z. Jiang, Z. Hao, M. C. Martin, P. Kim, H. L. Stormer, and D. N. Basov, Nature Phys. 4, 532 (2008).
\bibitem{Kosmider13}  K. Ko\'{s}mider, J. W. Gonz\'{a}lez, and J. Fern\'{a}ndez-Rossier, Phys. Rev. B 88, 245436 (2013).
\bibitem{Liu13}  G. -B. Liu, W.-Y. Shan, Y. Yao, W. Yao, and D. Xiao, Phys. Rev. B 88, 085433 (2013).
\bibitem{Kormanyos13}  A. Korm\'{a}nyos, V. Z\'{o}lyomi, N. D. Drummond, P. Rakyta, G. Burkard, and V. I. Fal'ko, Phys. Rev. B 88, 045416 (2013).
\bibitem{Cheiw12}  T. Cheiwchanchamnangij and W. R. L. Lambrecht, Phys. Rev. B 85, 205302 (2012).
\bibitem{Mak10} K. F. Mak, C. Lee, J. Hone, J. Shan, and T. F. Heinz, Phys. Rev. Lett. 105, 136805 (2010).
\bibitem{Sun14}  D. Sun, Y. Rao, G. A. Reider, G. Chen, Y. You, L. Br\'{e}zin, A. R. Harutyunyan, and T. F. Heinz, Nano Lett. 14, 5625 (2014).
\bibitem{Kumar14}  N. Kumar, Q. Cui, F. Ceballos, D. He, Y. Wang, and H. Zhao, Phys. Rev. B 89, 125427 (2014).


\end{thebibliography}
\end{document}